\documentclass{aa}  

\usepackage{graphicx}
\usepackage{color}
\usepackage{footnote}
\makesavenoteenv{tabular}
\usepackage{wasysym}
\usepackage{amsmath}

\usepackage{newtxtext} 
\usepackage{newtxmath} 
\usepackage{orcidlink} 
\usepackage{siunitx}
\usepackage{upgreek}

\usepackage{hyperref}

\hypersetup{
    colorlinks,
    linkcolor={red!50!black},
    citecolor={blue!50!black},
    urlcolor={blue!80!black},
    pdftitle={Potential of VLTI observations for circumstellar disk variability},
    pdfauthor={A. Bensberg, J. Kobus, S. Wolf},
    pdfkeywords={radiative transfer, methods: numerical,  circumstellar matter, protoplanetary disks, stars: variables,  ISM: clouds
    }
}

\DeclareMathAlphabet\mathcal{OMS}{txsy}{m}{n}
\DeclareSIUnit\jy{Jy}
\DeclareSIUnit\au{au}
\DeclareSIUnit\msun{M\ensuremath{_{\odot}}}
\DeclareSIUnit\lsun{L\ensuremath{_{\odot}}}
\DeclareSIUnit\pc{pc}

\begin{document} 

    \title{The potential of VLTI observations for the study of circumstellar disk variability}
    
    \author{A. Bensberg \orcidlink{0000-0001-6789-0296}, J. Kobus \orcidlink{0000-0002-3741-5950} and S. Wolf \orcidlink{0000-0001-7841-3452}}
    
    \institute{Institut für Theoretische Physik und Astrophysik, Christian-Albrechts-Universität zu Kiel, Leibnizstr. 15, 24118 Kiel, Germany\\
    \email{\href{mailto:abensberg@astrophysik.uni-kiel.de}{abensberg@astrophysik.uni-kiel.de}}}
    
    \date{Received / Accepted }
    
    \abstract
    {A characteristic feature of young stellar objects is their variability, which is caused by a variety of different physical processes. High-resolution interferometric observations in the near- and mid-infrared wavelength ranges spanning multiple epochs allow the detailed study of these processes.}
    {We aim at investigating the expected variations of the interferometric observables connected to changes in the measured photometric fluxes of a typical variable accreting central young stellar object with a circumstellar disk.}
    {We calculated visibilities and closure phases as well as the photometric flux of brightness distributions obtained using 3D Monte Carlo radiative transfer simulations for a model of a circumstellar disk with an accreting central star.}
    {Changes in the accretion luminosity of the central object, that is, an accreting pre-main-sequence  star, can lead to significant variations in the visibility and closure phase of the star--disk system measured with instruments at the Very Large Telescope Interferometer (VLTI) that can be related to changes in the photometric flux. Taking into account additional effects due to baseline variation, interferometric observations can provide valuable contributions to the understanding of the underlying processes. Additionally, we provide the web application \texttt{VLTI B-VAR} that allows  the impact of the hour angle on the visibility and closure phase for customized intensity maps to be estimated.}
    {}
    
    
    \titlerunning{VLTI observations of disk variability}
    \maketitle

\section{Introduction}\label{sec:Intro}
The evolution of circumstellar disks is a highly dynamic process. To understand the underlying physical mechanisms, it is vital to study disks at different evolutionary stages. However, individual observations and most observational projects  only provide snapshots of a certain evolutionary stage, even though variability is known to be a vital feature in such objects on various timescales.
Strong photometric variability was measured for 11 T Tauri-like objects by \cite{Joy1945}. With an increasing number of variable sources found over the next decades, it is currently estimated, based on observations performed with the Spitzer Space Telescope, that roughly half of all pre-main-sequence stars show variability in the infrared on various timescales \citep[e.g.,][]{Rebull2015,Poppenhaeger2015,Wolk2018}.

In this paper, we examine the potential of using interferometric measurements to study the temporal variability of young star-disk systems. Photometric variability has been studied for decades and allows   conclusions to be drawn about the overall strength and duration of the corresponding events \citep[e.g.,][]{Stauffer2016,Pouilly2021,Cody2022}. However, the restriction to the total flux of the observed systems allows to derive only weak constraints regarding the underlying variability of the spatial structure of these systems. This deficiency can be overcome by high-resolution interferometric observations that are sensitive to variations in the brightness distribution of the circumstellar material. This kind of observations can be performed by using modern interferometric instruments such as MATISSE, GRAVITY, or PIONIER available at the Very Large Telescope Interferometer (VLTI). The potential of such high-resolution multi-epoch observations is demonstrated in the first studies, where the variability of the innermost region of circumstellar environments is investigated \citep[e.g.,][]{Kluska2016, Chen2018, Chen2019}. Moreover, \cite{Kobus2020} show that VLTI observations of selected protoplanetary disks performed at multiple epochs contain evidence for variability of various sources. As multi-epoch observations are rather time-consuming, corresponding observations have to be prepared carefully.

To provide a basis for the preparation of future interferometric variability studies, we present the expected variations in the interferometric quantities of a basic protoplanetary disk model resulting from temporal variations in the accretion luminosity.
This is done by deriving synthetic interferometric observations for brightness distributions of a circumstellar disk with a central young stellar object obtained with Monte Carlo radiative transfer simulations. As the aim is to  observe the inner regions of circumstellar disks using the VLTI, we chose a wavelength range covering the $H$ band (PIONIER) and the $K$ band (GRAVITY), as well as the $L$, $M$, and $N$ bands (MATISSE). Thus, both the thermal reemission radiation of the warm dust and the direct and scattered stellar radiation were considered.

In Sect.~\ref{sec:Meth} we describe the procedure applied to get from the circumstellar disk model to synthetic observations.
The influence of temporal variations of a model with variable accretion luminosity of the central star on the visibility and closure phase is presented in Sect.~\ref{sec:Res}. An increasing stellar luminosity leads to increased heating of the dust in the disk, resulting in an increase in the inner radius of the disk due to sublimation. Additionally, the illumination of the disk and the corresponding temperature distribution changes. To isolate these two effects, we first study the influence of a geometrical change due to an increasing inner radius in Sect.~\ref{subsec:RadRat}. The impact of an increase in stellar luminosity with a constant inner radius is then discussed in Sect.~\ref{subsec:FluxRat}. Finally, we consider both effects using a model with increasing stellar luminosity and dust sublimation in Sect.~\ref{subsec:SubRad}.
Lastly, we discuss the observational consequences of our findings, as well as other effects on the observables, and give a brief outlook on future studies in Sect.~\ref{sec:Disc}.

\section{Method}\label{sec:Meth}

Synthetic observations were used to quantify the impact of variability on the visibility and closure phase for the different model setups. We calculated brightness distributions using radiative transfer simulations for a basic, often applied circumstellar disk model. In the following we outline the underlying model as well as the radiative transfer simulations.

\paragraph{Model: Circumstellar disk}
We chose a general circumstellar disk model that had been successfully used in multiple previous studies of protoplanetary disk properties (e.g.~\citealp{Sauter2009};~\citealp{Brunngraeber2016};~\citealp{Hofmann2022}). The corresponding density distribution of the dust is based on the disk models of~\cite{ShakuraSunjaev}. Using the cylindrical coordinates $(r,z)$ it is given by
\begin{equation}
    \label{eq:dens_dist}
    \rho (r,z) = \rho_0 \left(\frac{r}{R_0}\right)^{-\alpha} \exp\left[ - \frac{1}{2} \left(\frac{z}{h(r)}\right)^2\right].
\end{equation}
Here $\rho_0$ denotes the density scaling parameter determined by the disk mass, $\alpha$ the radial density exponent, and $h(r)$ the scale height of the disk depending on the reference scale height $h_{\rm ref}$ at the reference radius $R_0$ and the flaring parameter $\beta$:
\begin{equation}
    \label{eq:scale_height}
    h(r) = h_{\rm ref} \left(\frac{r}{R_0}\right)^\beta.
\end{equation}
For the disk parameters we chose values of $\alpha = 1.8$~\citep{Andrews2010}, $\beta = 1.1$~\citep{Woitke2019}, $R_0 = \SI{100}{\au}$, and $h_{\rm ref} = \SI{10}{\au}$. The density distribution of the dust was then calculated between a given inner radius $R_{\rm in}$ and the outer radius $R_{\rm out}$, which  in this case was identical to the reference radius.

\paragraph{Model: Central star}
The parameter of the embedded young stellar object (hereafter central star) were varied for the different cases. However, the reference model for the central star was defined by a luminosity of $L_{\star} = \SI{1.5}{\lsun}$ and an effective temperature of $T_{\star} = \SI{4000}{\K}$, typical of a low-mass TTauri star.
In order to account for the change in the stellar spectral energy distribution during an accretion event, all simulations were also performed for a stellar effective temperature of \SI{8000}{\K}, which had already been successfully used to model variable stars during an accretion event in circumstellar disks \citep[e.g.,][]{Schegerer2009,Brunngraeber2016}.
In the following, we   refer to simulations with an effective temperature of $T_{\star} = \SI{4000}{\K}$ as the ``ground state'' and simulations with an effective temperature of $T_{\star} = \SI{8000}{\K}$ as the ``active state''. 

\paragraph{Model: Dust properties}
We assumed that the dust composition is similar to that of the ISM and thus applied a mixture of 62.5\% astronomical silicate and 37.5\% graphite. The graphite component followed the typical $\frac{1}{3}$-$\frac{2}{3}$ relation for parallel and perpendicular orientations by \cite{DraineMalhotra1993}. The grain size distribution followed the power law of \cite{MathisRumplNordsieck1977} with $n(s) \propto s^{-q}$ for $q=3.5$ and grain radii between \SI{5}{\nm} and \SI{250}{\nm}. This size distribution was consistent with the observational constraints for the dust in upper disk layers found in the near- to mid-infrared wavelength range considered in our study. The optical properties of the compact chemically homogeneous dust grains were derived on the basis of the Mie scattering theory using the wavelength-dependent refractive indices of \cite{Draine1984}, \cite{Laor1993}, and \cite{Weingartner2001}.

\paragraph{Radiative transfer}
Based on the above model, we simulated brightness distributions using the publicly available 3D Monte Carlo radiative transfer code \texttt{POLARIS} \citep{Reissl2016}.\footnote{\url{https://portia.astrophysik.uni-kiel.de/polaris/}} \texttt{POLARIS} makes use of the immediate correction method of~\cite{BjorkmanWood2001} and the instant reemission method of~\cite{Lucy1999} for the simulation of temperature distributions. The thermal reemission radiation of the dust was calculated using a ray-tracing algorithm, while the scattered stellar light was determined using an anisotropic Monte Carlo scattering method.

\paragraph{Interferometric observables}
Visibilities and closure phases of the resulting ideal intensity maps were then calculated using fast Fourier transforms. The uv coverage was chosen according to the baselines covered by the VLTI. For this we used two approaches. In the case of the investigation of the impact of variations in the inner rim, the stellar luminosity, and the combination of both on the visibilities (Sections \ref{subsec:RadRat}, \ref{subsec:FluxRat}, and \ref{subsec:SubRad}, respectively), we considered one baseline with a length of \SI{130}{\meter}. This corresponds to the longest baselines of the “UT” configuration and the   “Large” configuration.\footnote{ESO - VLTI Configuration Overview - P112\\\url{https://www.eso.org/sci/facilities/paranal/telescopes/vlti/configuration/P112.html}} In the case of models with inclined disks, we considered two baseline orientations, one parallel (major) and one perpendicular (minor) to the axis of rotation of the disk. For non-inclined, and thus radially symmetric disks, the visibility was independent of the orientation of the baseline.

In the second approach, that is, when we calculated closure phases or when we studied the impact of the Earth's rotation on the projected baseline and the resulting influences on visibilities and phases (Section \ref{subsec:basvar}), we used the uv coverage of an exemplary object with a declination of \SI{-5}{\degree} (e.g., Orion Nebula) that would be obtained when using the UT configuration at an hour angle of \SI{0}{\hour}. Unless explicitly stated otherwise, we show the closure phases for the triangle with the longest baselines.

Following \cite{Lopez2022}, we assumed that the detection limits for the visibility and closure phases are 0.1 and \ang{5}, respectively.
The distance to the circumstellar disk model was set to \SI{140}{\pc}, corresponding to the distance to the Taurus molecular cloud.

\paragraph{Parameter space}
Since this study focuses on a model of variability caused by accretion, we focused on the accretion luminosity. Within the simulations, the accretion and stellar luminosity were combined into one central radiation source. Therefore, in the following, stellar luminosity is used as a term for the luminosity of the central source including both components.

Bright outburst events in circumstellar disks can shift the sublimation radius outward, and thus can increase the inner radius of the disks. This leads to a different illumination, and thus heating and light scattering by the disk. It will therefore affect the measured visibilities and closure phases. Thus, we performed simulations of the disk models for different inner radii, ranging from \SI{0.1}{\au} to \SI{1.5}{\au}, corresponding to stellar luminosities ranging from \SI{1.5}{\lsun} to \SI{600}{\lsun}. 
The reference model has an inner radius of $R_{\rm in} = \SI{0.1}{\au}$ and a dust mass of \SI{e-5}{\msun} for all simulations.
A change in the inner radius of the disk was modeled by cutting off the inner parts of the density distribution up to the new radius. Thus, the disk mass decreases with increasing radii, while the local density in the disk remains unchanged.

It should be noted that the shape of the inner rim in our model is a vertical wall at the inner radius. While this is not in full agreement with the expected shape of an inner rim originating from hydrodynamical and dust sublimation processes \citep[e.g.,][]{Flock2016}, it will not significantly influence the general outcomes for the studied cases (see Appendix~\ref{sec:ModRin}) and is thus used for the sake of simplicity. A more detailed discussion of different inner rim shapes in the context of MATISSE observations and radiative transfer modeling can be found in \cite{Hofmann2022}.

The wavelengths used for all simulations correspond to the central wavelength of the bands covered by the beam combiners PIONIER, GRAVITY, and MATISSE, currently available at the VLTI: \SI{1.5}{\micro m} ($H$ band), \SI{2.2}{\micro m} ($K$ band), \SI{3.5}{\micro m} ($L$ band), \SI{4.8}{\micro m} ($M$ band), and \SI{10}{\micro m} ($N$ band).

\section{Results}\label{sec:Res}
If the luminosity of the central star with a circumstellar disk is increased, for example by an accretion event, there are two different mechanisms that influence the observational appearance of the disk. First, there is the change in illumination due to the increased stellar luminosity and the change in the stellar spectrum, which can be represented by an increase in the effective temperature of the star. Second, there is the increasing inner disk radius due to dust sublimation. In order to study the effect of the two mechanisms on the change in the photometric luminosity and the interferometric visibility, we first consider each mechanism individually. Finally, we present simulations for models with increased stellar luminosity and corresponding inner radii of the disk caused by dust sublimation.
In Sect.~\ref{subsec:RadRat} we discuss the influence of geometric changes in the disk due to an increasing inner radius. The impact of changes in stellar luminosity and effective temperatures are outlined in Sect.~\ref{subsec:FluxRat}. The influence of the combined mechanisms on the photometric luminosity and interferometric visibility is shown in Sect.~\ref{subsec:SubRad}.

\subsection{Increasing inner radius}\label{subsec:RadRat}
\begin{figure}[!h]
    \centering
    \includegraphics[width=1.\linewidth]{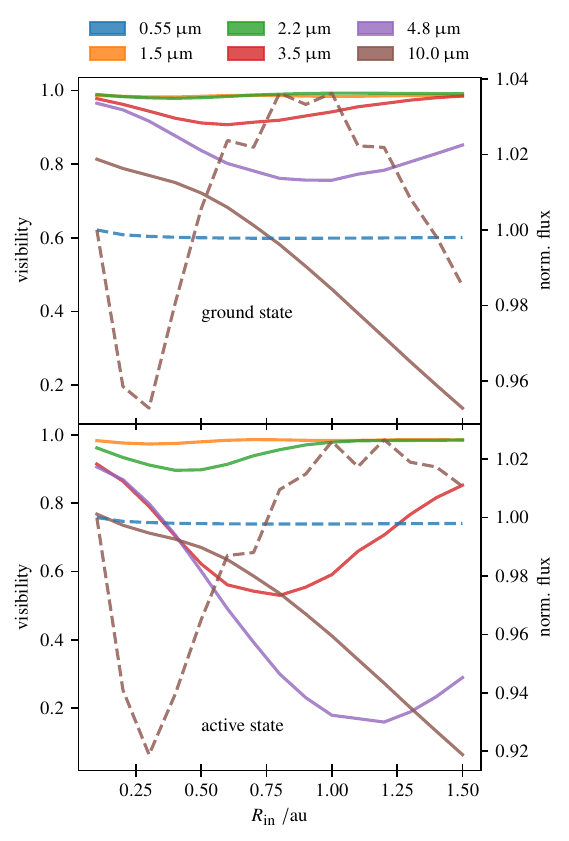}
    \caption{Visibilities for ground state (top) and active state (bottom) models as a function of the inner disk radius for the $H$, $K$, $L$, $M$, and $N$ bands (solid; left axis). The corresponding fluxes in the photometric $V$ and $N$ bands are also shown (dashed; right axis). The $V$- and $N$-band fluxes are normalized with respect to the reference model with $R_{\rm in}=\SI{0.1}{\au}$ (see Sect.~\ref{subsec:RadRat} for details).} 
    \label{fig:radratio}
\end{figure}

To isolate the effect of an increase in the inner disk radius, we first studied disk models with different inner radii and a fixed stellar luminosity of \SI{1.5}{\lsun}. The resulting images of the direct and scattered stellar radiation and reemitted thermal radiation were then used to calculate the visibility of interferometric observations with a baseline of \SI{130}{\m}, corresponding to the longest baseline of the VLTI.
The calculated visibilities as well as the photometric fluxes in the $V$ and $N$ bands for both the ground state and the active state are shown in Fig.~\ref{fig:radratio} for a disk seen face-on.

We find that a change in the inner radius has a minor impact on the visibilities at the $H$ and $K$ bands only. At these wavelengths, the intensity is dominated by the contribution of the star and the region near the inner rim, which are both unresolved. The visibilities of the $L$ and $M$ bands show a minimum for the inner radii at which the inner rim is resolved. The appearance of the disk in the $N$ band is mainly dominated by the reemitted thermal radiation of the disk. By increasing the inner radius of the disk, the reemitting surface is also increased, leading to a steady decrease in the visibility. Since even the largest inner rim within our parameter space is not resolved, the first minimum of the visibility distribution is not reached.

The photometric flux of the $V$ band is dominated by the direct contribution of the star and is, except for a small deviation caused by a change in the scattering probability at the smallest inner radii, not influenced by the change in the inner disk radius. In contrast to this, there is a change in the $N$-band flux up to \SI{4}{\percent} in both directions. While the flux first decreases because the hot inner disk region is cut off, it increases for inner disk radii larger than \SI{0.5}{\au}. This is caused by an increase in the emitting surface area due to an increased efficiency of disk illumination, and thus heating of regions that are shadowed by the optical thick inner disk regions at smaller inner disk radii. 
\begin{figure*}[!h]
    \centering
    \includegraphics[width=.9\linewidth]{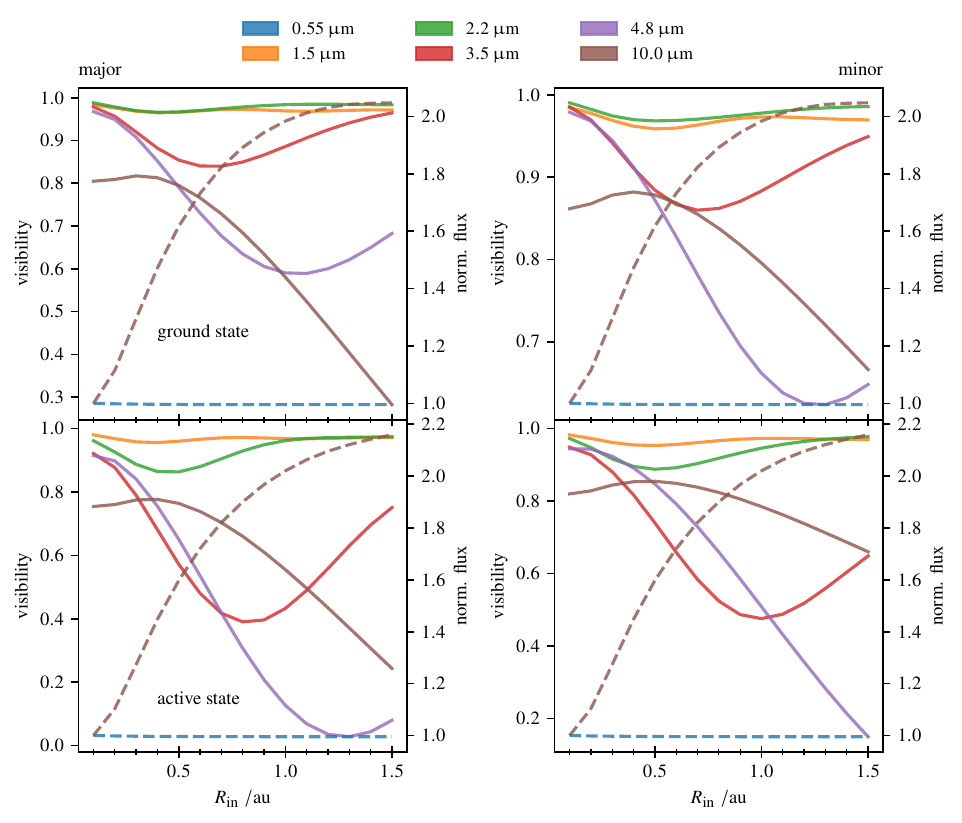}
    \caption{Visibilities for ground state (top) and active state (bottom) models with an inclination of \ang{45} as a function of the inner disk radius for the $H$, $K$, $L$, $M$, and $N$ bands (solid; left axis) for baseline orientations along the major (left) and minor axis (right) of the disk image. The corresponding fluxes in the photometric $V$ and $N$ bands are also shown (dashed; right axis).  The $V$- and $N$-band fluxes are normalized with respect to the reference model with  $R_{\rm in}=\SI{0.1}{\au}$ (see Sect.~\ref{subsec:RadRat} for details).}
    \label{fig:radratio45}
\end{figure*}
However,  the $N$-band flux  decreases again as the inner disk radius is further increased, cutting off regions containing warm dust emitting in the $N$ band.

The simulations of the active state follow the above trends. However, the changes in the visibilities and $N$-band photometry are larger than for the ground state. This can be explained by the more efficient heating of the disks dust due to an increased absorption efficiency at the shorter wavelengths of the stellar emission.

Due to the asymmetry of the image of an inclined disk, it is necessary to study those cases for different orientations of the baseline. The visibilities of the model described above with an inclination of \ang{45} are thus calculated for configurations orientated along the major and minor axis of the elliptical shape of the projected disk midplane.\footnote{The iso-intensity contours of the projected disk image have a non-complex structure, which is due to the vertical extension of the disk, the disk flaring, and the assumed anisotropic scattering function.} The corresponding plots of the visibilities as well as the photometric fluxes in the $V$  and $N$ bands for both stellar effective temperatures are shown in Fig.~\ref{fig:radratio45}.

While the trends found in the case of the disk in face-on orientation in Fig.~\ref{fig:radratio} are also shown here, the total change in the visibilities is larger. This is caused by the bright inner rim, which is directly in the line of sight of the observer due to the disk inclination. The influence of this bright structure gets larger for larger radii. This can also be seen from in the photometric $N$-band flux. The characteristic S-shape seen in Fig.~\ref{fig:radratio} is no longer present, and the total change in the flux is up to a factor of 2. The bright inner rim now dominates the $N$-band flux.

\begin{figure}[!h]
    \centering
    \includegraphics[width=1.\linewidth]{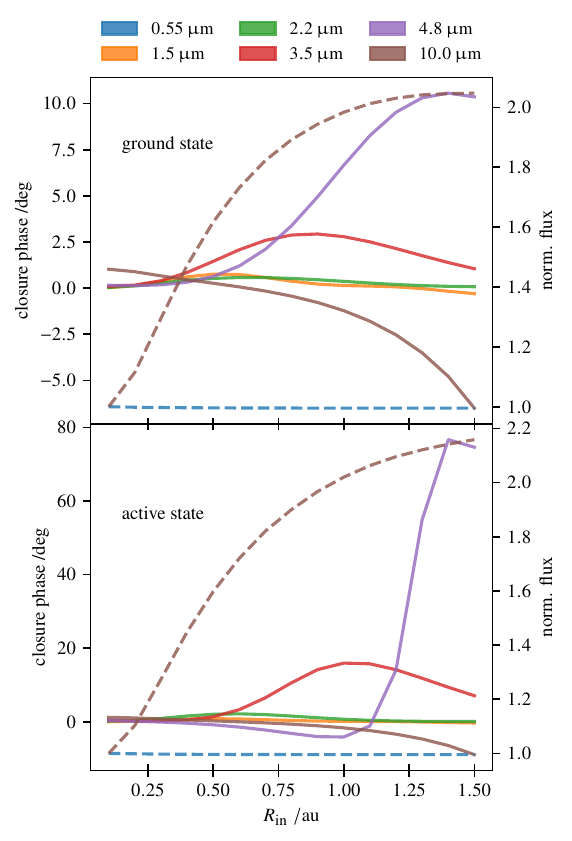}
    \caption{Closure phases for ground state (top) and the active state (bottom) models with an inclination of \ang{45} as a function of the inner disk radius for the $H$, $K$, $L$, $M$, and $N$ bands (solid; left axis). The corresponding fluxes in the photometric $V$ and $N$ bands are also shown (dashed; right axis). The $V$- and $N$-band fluxes are normalized with respect to the reference model with  $R_{\rm in}=\SI{0.1}{\au}$ (see Sect.~\ref{subsec:RadRat} for details).}
    \label{fig:radratiocp}
\end{figure}

In order to investigate whether accretion events lead to measurable symmetry changes, the closure phases are also calculated. 
Since the image of the disk seen in face-on orientation is point symmetric, only the case of inclined disks are considered in the following. The closure phases of the models with different inner radii for the configuration of the VLTI unit telescopes (UTs) with the largest baseline (i.e. the UT1-UT2-UT4 configuration) can be seen in Fig.~\ref{fig:radratiocp}.

Significant changes are only found for the $M$ and $N$ bands at larger inner radii. In the active state, the change in the closure phase for the $L$ band is also large enough to be measured. This is again caused by the bright extended inner rim that has a higher dust temperature for the case of an accretion event in our model.

In summary, we could show that the expected changes in the visibilities and closure phases caused by a change in the inner radius are significant if the inner radius gets large enough to be at least partially resolved. Furthermore, the flux of the disk must contribute significantly, which is the case at the longer wavelengths. Thus, this has to be taken into account when investigating variability mechanisms that include changes in the inner radius. However, within the considered parameter space, the inner radius has nearly no impact on the $V$-band photometry and will only lead to large changes in the $N$-band photometry in the case of an inclined disk.

\subsection{Increasing luminosity}\label{subsec:FluxRat}
\begin{figure}[!t]
    \centering
    \includegraphics[width=.985\linewidth]{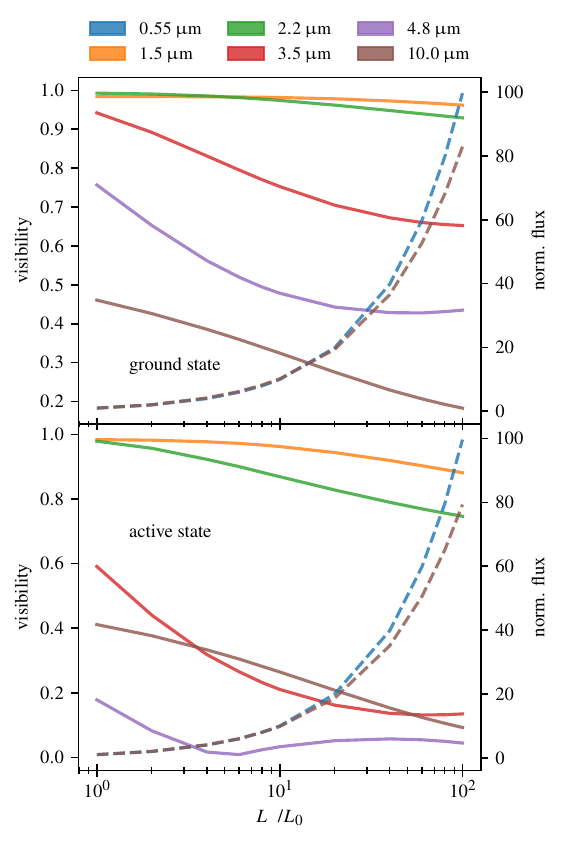}
    \caption{Visibilities for ground state (top) and the active state (bottom) models as a function of  stellar luminosity for the $H$, $K$, $L$, $M$, and $N$ bands (solid; left axis). The corresponding fluxes in the photometric $V$ and $N$ bands are also shown (dashed; right axis). The $V$- and $N$-band fluxes are normalized with respect to the reference model with $L = L_0$. The inner radius of the disk is set to $R_{\rm in}=\SI{1}{\au}$ for all models (see Sect.~\ref{subsec:FluxRat} for details).}
    \label{fig:fluxrat}
\end{figure}

In the following, we investigate the change in the brightness distribution due to a change in the luminosity of the central star. To do this, we perform stationary simulations of the model described in Sect.~\ref{sec:Meth} with fixed disk parameters and increase the luminosity of the star by up to a factor of 100. The inner radius of the disk model is set to \SI{1}{\au}, which is larger than the sublimation radius expected for the simulation with the highest luminosity and effective temperature to rule out dust temperatures beyond the sublimation temperature. The simulations are again performed for the ground state and the active state. The visibilities for all observing wavelengths are then calculated again using a configuration with a baseline of \SI{130}{\m}. 
The corresponding calculated visibilities as well as the photometric fluxes of the $V$ and $N$ bands for both stellar effective temperatures for a disk in face-on orientation can be found in Fig.~\ref{fig:fluxrat}.

In the case of the simulations for the ground state, the changes with increasing luminosity of the visibility of the $H$ and $K$ bands are well below 0.1. In contrast, the visibilities calculated for the $L$, $M$, and $N$ bands decrease significantly with increasing stellar luminosity. The results for the $H$ and $K$ bands can be explained by the dominance of the direct and scattered stellar radiation. Moreover, since the radial distribution of the scattered light does not significantly change at these wavelengths, the change in the visibility is insignificant, too. In the case of the longer wavelengths ($L$, $M$, $N$ bands), the contribution of the thermal reemission radiation of the dust to the net flux  increases significantly. An increase in the stellar luminosity leads to a stronger heating of the disk. Thus, the radial extent of the brightness distribution of the disk increases, which is also reflected in the corresponding visibilities.

\begin{figure*}[!t]
    \centering
    \includegraphics[width=.9\linewidth]{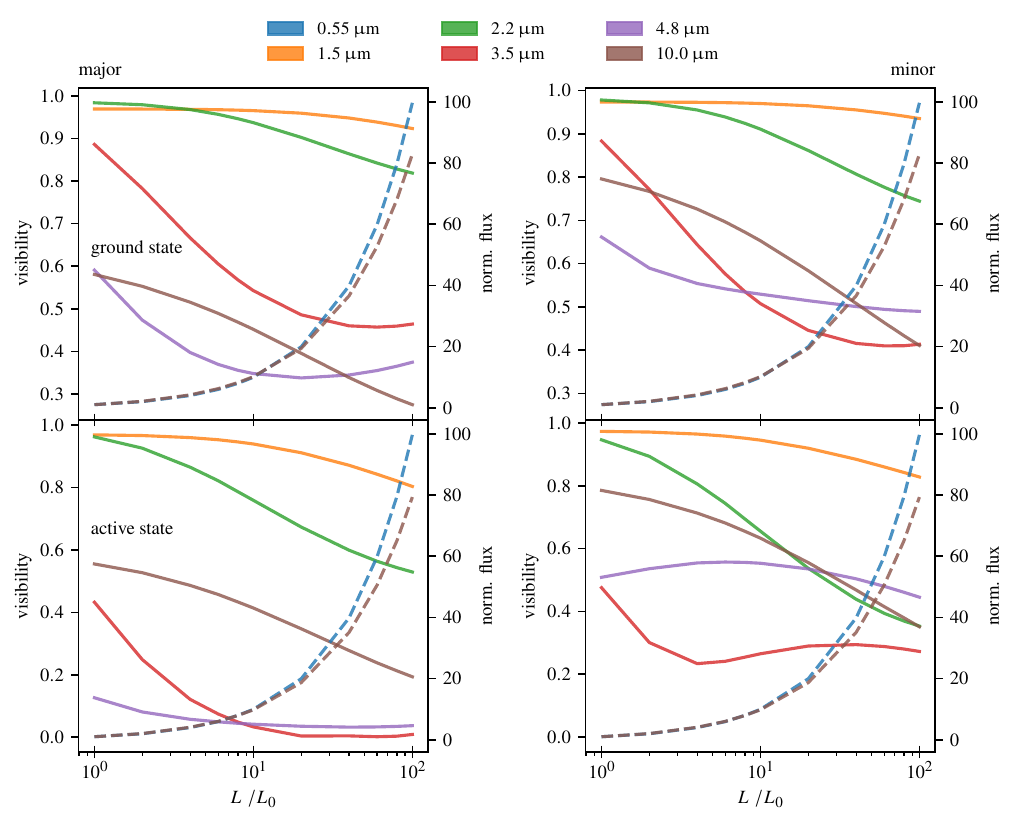}
    \caption{Visibilities for ground state (top) and active state (bottom) models with an inclination of \ang{45} as a function of stellar luminosity for the $H$, $K$, $L$, $M$, and $N$ bands (solid; left axis) for baseline orientations along the major (left) and minor axis (right) of the disk image. The corresponding fluxes in the photometric $V$ and $N$ bands are also shown (dashed; right axis). The $V$- and $N$-band fluxes are normalized with respect to the reference model with $L = L_0$. The inner radius of the disk is set to $R_{\rm in}=\SI{1}{\au}$ for all models (see Sect.~\ref{subsec:FluxRat} for details).}
    \label{fig:fluxrat45}
\end{figure*}
The photometric fluxes in the $V$ and $N$ bands increase with increasing stellar luminosity. While the $V$-band flux originates from the direct stellar radiation, and  thus follows the increase in total stellar luminosity, the $N$-band flux is mainly reemission radiation of the warm dust of the disk. Even though the heating is also more efficient at higher stellar luminosities, the $N$-band flux increases less than the $V$-band flux.

In the active state, the visibilities of the $H$, $K$, $L$, and $M$ bands  decrease with increasing stellar luminosity, and are lower than before. The shift of the stellar spectrum toward shorter wavelengths leads to a more efficient heating of the dust of the disk. Accordingly, the higher dust temperatures result in a larger amount of thermal reemission radiation in regions outside the innermost parts of the disk. This spatial broadening of the emitting regions of the disk cause the lower visibilities at these wavelengths. The change in the visibility in the $N$ band in the active state is similar to the change in the ground state since the corresponding emitting outer disk regions containing warm dust are affected only marginally by the change in the stellar effective temperature.

This also applies to the $N$-band photometric flux. The $V$-band flux again follows the increase in stellar luminosity.
As in Sect.~\ref{subsec:RadRat}, the simulations were also done for the case of a disk with an inclination of \ang{45}. The resulting visibilities and baseline orientations are shown in Fig.~\ref{fig:fluxrat45}.
It can be seen that the visibilities of the $K$, $L$, and $M$ bands for the orientation along the major axis are lower than in the face-on case. This is caused by an increase in the flux per area of the disk, while the unresolved direct stellar radiation does not change. This effect is even stronger in the $H$ and $K$ bands at the highest stellar luminosity, where parts of the hot inner rim contribute to the resolved flux.
While the shape of the visibility distributions of the $H$, $K$, $L$, and $M$ bands are mainly determined by the ratio of unresolved to resolved flux, the shape of the visibility distribution of the $N$ band is determined by the increase in the heated disk area in the line of sight. The direct stellar contribution to the $N$-band visibilities is three orders of magnitude smaller than that of the disk, and is thus negligible.

The visibilities calculated using the orientation along the minor axis are higher than those calculated for the orientation along the major axis because of a narrow brigthness distribution. In the case of the $H$ and $K$ bands the hot inner rim contributes to the resolved flux, and thus decreases the star-to-disk flux ratio, leading to lower visibilities.

\begin{figure}[!ht]
    \centering
    \includegraphics[width=1.\linewidth]{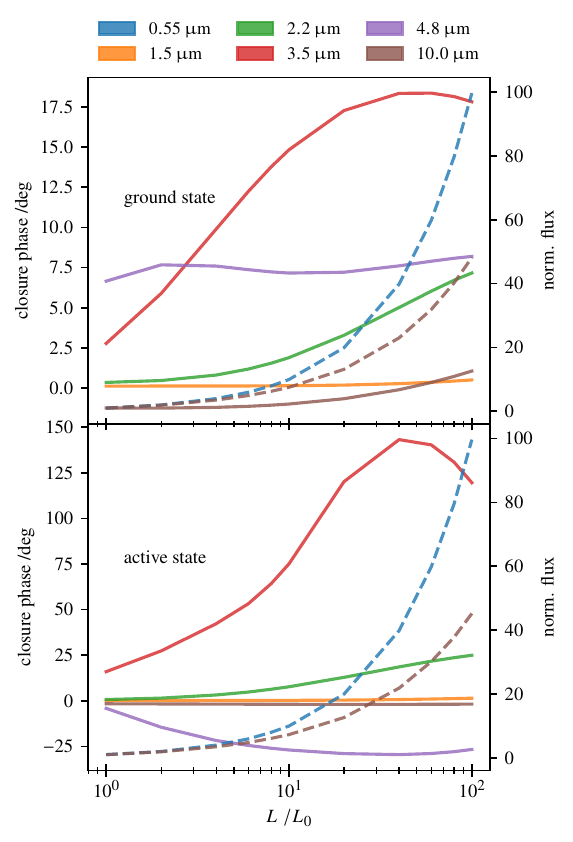}
    \caption{Closure phases for ground state (top) and the active state (bottom) models with an inclination of \ang{45} as a function of  stellar luminosity for the $H$, $K$, $L$, $M$, and $N$ bands (solid; left axis). The corresponding fluxes  in the photometric $V$ and $N$ bands are also shown (dashed; right axis). The $V$- and $N$-band fluxes are normalized with respect to the reference model with $L = L_0$. The inner radius of the disk is set to $R_{\rm in}=\SI{1}{\au}$ for all models (see Sect.~\ref{subsec:FluxRat} for details).}
    \label{fig:fluxratcp}
\end{figure}

The simulations for the inclined disk are then again used to calculate the closure phases for the UT configuration. The resulting closure phases for the triangle with the longest baselines are shown in Fig.~\ref{fig:fluxratcp}.
It can be seen that the changes in the closure phase of the $L$ band are the largest. Since the $L$-band flux of the disk originates mainly at the hot inner rim, the symmetry of the corresponding spatial brightness distribution strongly depends on the heating of the inner rim that is stronger in the active state. While this is also true for the $H$ and $K$ bands, the emission at these wavelengths is dominated by the direct emission of the star. The change in the closure phases is therefore smaller. The $N$-band flux originates in regions of warm dust farther out in the disk. Even though the total flux from these regions increases with increasing stellar luminosity, the symmetries of the emitting regions, and thus the closure phases, do not change drastically.

In summary, an increase in the stellar luminosity has a large effect on the visibilities of the $L$, $M$, and $N$ bands. Changes in the visibilities of the shorter wavelengths are only significant in the case of an accretion event or an inclined disk. The influence on the closure phase is also largest for the $L$ band. However, significant changes can also be found for the $K$ band and for the case of an accretion event for every wavelength except for the $N$ band.

\subsection{Increasing luminosity with sublimation}\label{subsec:SubRad}
\begin{figure}[!t]
    \centering
    \includegraphics[width=1.\linewidth]{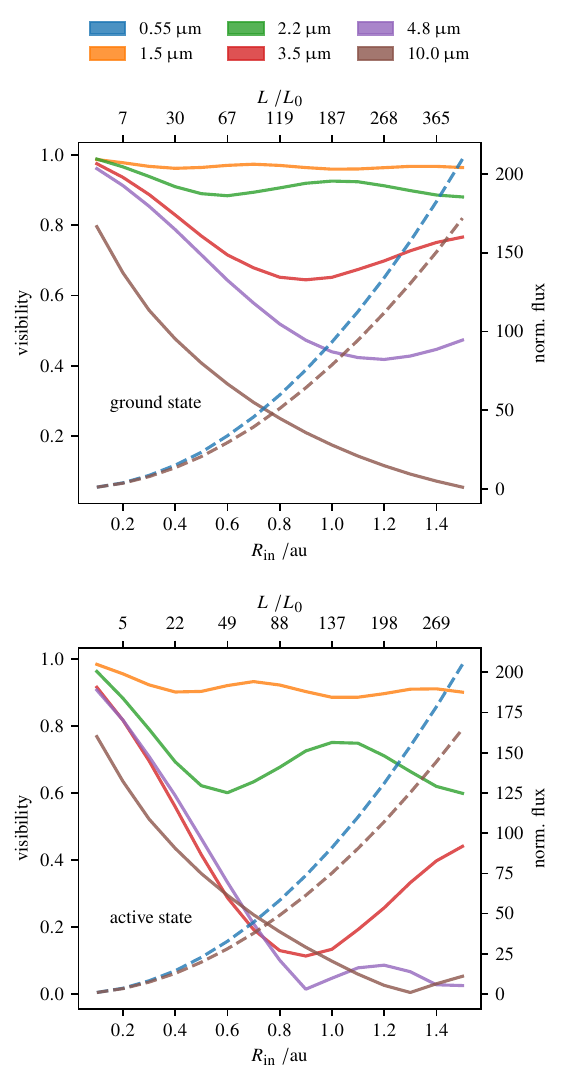}
    \caption{Visibilities for ground state (top) and the active state (bottom) models as a function of   stellar luminosity, and the corresponding sublimation radius for the $H$, $K$, $L$, $M$, and $N$ bands (solid; left axis). The corresponding fluxes in the photometric $V$ and $N$ bands are also shown (dashed; right axis). The $V$- and $N$-band fluxes are normalized with respect to the reference model with $L = L_0$. The stellar luminosity of each model is increased such that the sublimation radius matches the given $R_{\textrm{in}}$ (see Sect.~\ref{subsec:SubRad} for details).}
    \label{fig:subratio}
\end{figure}
\begin{figure*}[!t]
    \centering
    \includegraphics[width=.85\linewidth]{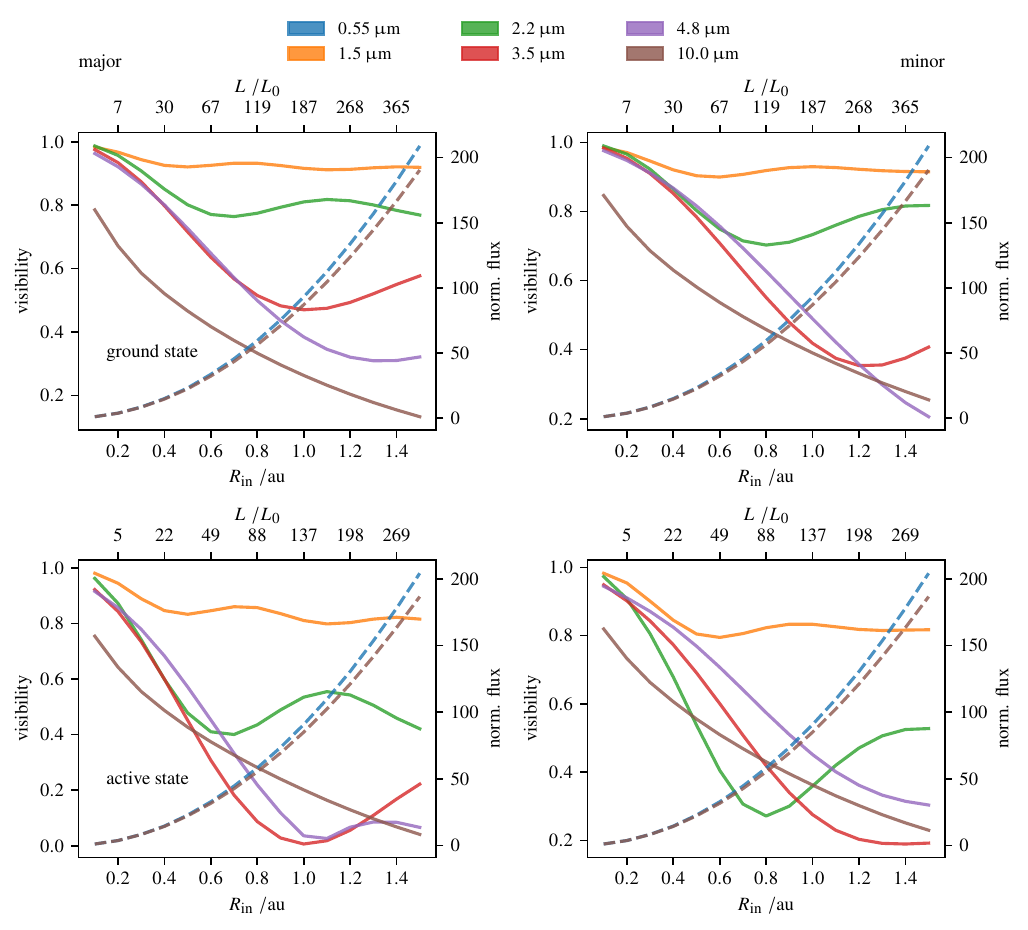}
    \caption{Visibilities for ground state (top) and the active state (bottom) models with an inclination of \ang{45} as a  function of   stellar luminosity and the corresponding sublimation radius for the $H$, $K$, $L$, $M$, and $N$ bands (solid; left axis) for baseline orientations along the major (left) and minor axis (right) of the disk image. The corresponding fluxes in the photometric $V$ and $N$ bands are also shown (dashed; right axis). The $V$- and $N$-band fluxes are normalized with respect to the reference model with $L = L_0$. The stellar luminosity of each model is increased such that the sublimation radius matches the given $R_{\textrm{in}}$ (see Sect.~\ref{subsec:SubRad} for details).}
    \label{fig:subratio45}
\end{figure*}
Finally, we discuss the combination of the two effects described above, an increasing inner radius due to sublimation caused by an increasing stellar luminosity. As in Sect.~\ref{subsec:RadRat}, we simulated models with inner radii between \SI{0.1}{\au} and \SI{1.5}{\au}. The stellar luminosity was then increased to match the estimated sublimation radius with the inner radius. In order to model an accretion event, the simulations were again done for the ground state and the active state. We then calculated the visibility of the intensity distributions for the configuration with the longest baseline,  \SI{130}{\m}. The resulting visibilities as well as photometric fluxes of the $V$ and $N$ bands are shown in Fig.~\ref{fig:subratio}.

In the ground state, the change in the visibility with increasing sublimation radius is measurable for all wavelengths except for the $H$ band. Similar to the cases discussed in Sects.~\ref{subsec:FluxRat} and~\ref{subsec:RadRat}, the flux at \SI{1.5}{\micro\meter} is mainly dominated by the stellar emission. 
Thus, the star-to-disk flux ratio does not significantly change in the $H$ band when increasing the stellar luminosity and subsequently the inner radius, leading to only small changes in the visibilities. In comparison with the simulations with fixed luminosity in  Fig.~\ref{fig:radratio} in Sect.~\ref{subsec:RadRat} and with fixed inner radius in  Fig.~\ref{fig:fluxrat} in Sect.~\ref{subsec:FluxRat}, the change in visibility is now larger than for the single effects alone. In particular, the $N$-band visibilities now rapidly decline, caused by a combination of the effects of better illumination due to the geometric change in and stronger heating of the dust due to a higher stellar luminosity.

In the active state, the changes in the visibilities are even stronger, with the visibilities of the $L$ and $N$ bands getting close to and the $M$ band even reaching zero.
This is caused by an increase in the disk-to-star flux ratio over all wavelengths. It can also be seen that structures of the disk can be resolved for every band except the $N$ band.

Similar strong effects can also be seen for the simulations of the inclined disk. The corresponding visibilities are shown in Fig.~\ref{fig:subratio45}.
In the ground state with the baseline orientation along the major axis, the visibilities of the $K$, $L$, and $M$ bands are smaller than for the face-on simulations, caused by the changed disk-to-star ratio. Changes in the visibilities for the $H$ band are larger than 0.1, and are thus measurable.

These effects are even stronger in the active state. The changes in the visibilities are now up to 0.2, and the $K$-band visibilities are at the level of those of the $L$ band for the face-on simulations. In the case of the $M$ band, the hot inner rim contributes to the unresolved flux, leading to a small increase in the visibilities to values above zero. This can also be seen for the $N$-band visibilities.
The minima of the visibilities of the simulations with the baseline orientated along the minor axis of the disk are shifted toward larger radii. This is caused by the inner rim appearing smaller due to the inclination.
In the active state, the hot inner rim   contributes to the resolved flux of the $M$-band simulation, strongly decreasing the corresponding visibilities.

\begin{figure}[!t]
    \centering
    \includegraphics[width=1.\linewidth]{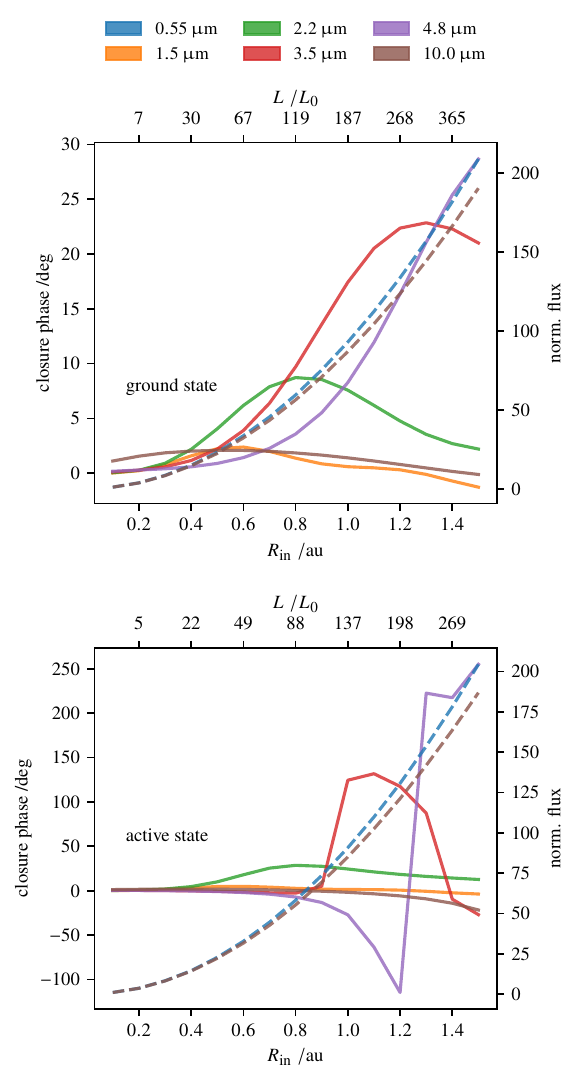}
    \caption{Closure phases for ground state (top) and active state (bottom) models with inclination of \ang{45} as a function of  stellar luminosity and the corresponding sublimation radius for the $H$, $K$, $L$, $M$, and $N$ bands (solid; left axis). The corresponding fluxes in the photometric $V$ and $N$ bands are also shown (dashed; right axis). The $V$- and $N$-band fluxes are normalized with respect to the reference model with $L = L_0$. The stellar luminosity of each model is increased such that the sublimation radius matches the given $R_{\textrm{in}}$ (see Sect.~\ref{subsec:SubRad} for details).}
    \label{fig:subratiocp}
\end{figure}

The corresponding closure phases calculated for the configuration with the longest baselines are shown in Fig.~\ref{fig:subratiocp}.
In comparison to Fig.~\ref{fig:fluxratcp} (in Sect.~\ref{subsec:FluxRat}), the $K$-band closure phases are larger than in the case with fixed inner radius. The hot inner rim contributes significantly to the $K$-band flux, but only if dust temperatures close to the sublimation temperature are reached. Since parts of the inner rim are in the line of sight of the observer in the case of an inclined disk, the asymmetry gets stronger, and thus the changes in the closure phase increase.

This effect is even stronger for the active state. The point at which the visibility of the $M$ band seen in Figs.~\ref{fig:subratio} and~\ref{fig:subratio45} reaches zero can now be recognized by a flip of the sign of the closure phase at $R_{\textrm{in}} = \SI{1.3}{\au}$.
The closure phase of the $L$-band simulations also show a strong increase to values up to \ang{150}. The peak of the distribution also corresponds to the minimum of the visibility for the $L$ band in Fig.~\ref{fig:subratio45}.

We conclude that the combination of the effects described in Sects.~\ref{subsec:RadRat} and~\ref{subsec:FluxRat} lead to measurable changes in the visibilities and closure phases for all studied wavelengths. To make sure that the simplification made for the shape of the inner rim of the disk does not affect the trends found above, we repeated the simulations with a model including a modified inner rim (see Appendix~\ref{sec:ModRin}). In summary, all trends and the magnitude of variations discussed above were also found for the model with modified inner rim.

\section{Discussion}\label{sec:Disc}
The simulations presented above showed significant changes in the visibilities and closure phases, and in the photometric fluxes. It could therefore be possible to observe variability due to accretion using interferometry if it has been measured photometrically, and vice versa. Interferometric observations could thus provide valuable information about the structure of these variable sources. However, the chosen model only allows inferring changes in the brightness distribution of the disk caused by an increase in the luminosity of the central source. Additional sources of variability like embedded companions \citep[e.g.,][]{Brunngraeber2018} or stellar spots were not considered and will change the brightness distribution and the corresponding observable. Nevertheless, the model provides an indication of the strength of the changes that can be expected for variable objects.

Considering the conditions for observations, it should be noted that the correlated fluxes calculated for all wavelengths of the model are well above the VLTI detection limits. The effects discussed in Sect.~\ref{sec:Res} are thus observable, in principle. However, another important aspect of planning observations or analyzing archive data is the baseline variation due to a change in the hour angle. Since the apparent changes in the visibilities and closure phases caused by this effect are on the same order as those due to illumination variation, we present a detailed discussion of the baseline variation in Sect.~\ref{subsec:basvar}. Corresponding considerations concerning the timescales of possible observations and our model simulations are discussed thereafter in Sect.~\ref{subsec:obscale}.

\subsection{Baseline variation}\label{subsec:basvar}
Temporal variability is a vital feature of circumstellar disks on various timescales, ranging from hours to years. When studying such variations by comparing multi-epoch observations, it should be ensured that the differences in the visibilities and closure phases caused by deviations of the baselines due to a different hour angle are smaller than the differences expected from the temporal variations of the brightness distribution. 
\begin{figure*}[!t]
    \centering
    \includegraphics[width=.85\linewidth]{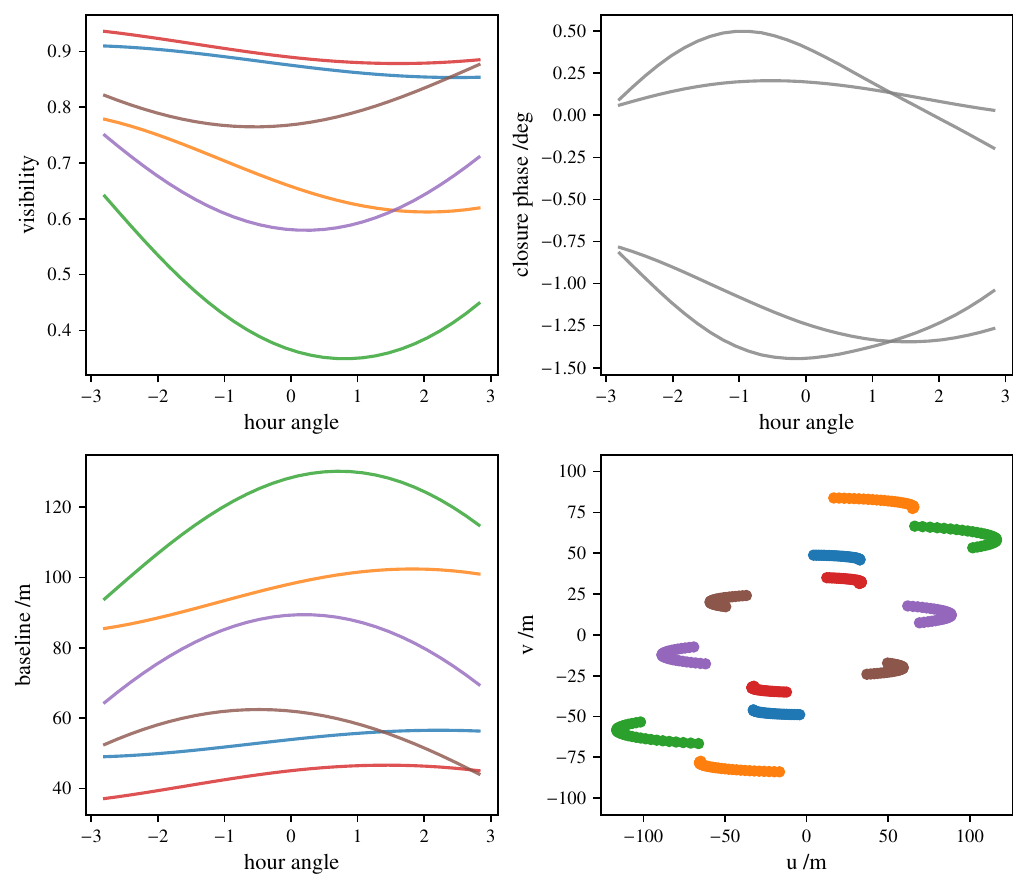}
    \caption{Visibility and closure phase as a function of hour angle for the active state model with   inner radius of $R_{\textrm{in}} = \SI{0.5}{\au}$, inclination of \ang{45}, as well as dust sublimation (see Sect.~\ref{subsec:SubRad} for details), for the $L$ band (top). The change in the length of the baselines used to calculate the visibilities and the corresponding uv coverage (bottom) are shown in their associated colors.}
    \label{fig:blvar_eg}
\end{figure*}
In order to distinguish these phenomena, we calculate the visibilities and closure phases of the disk model with dust sublimation discussed in Sect.~\ref{subsec:SubRad} for different hour angles. Therefore, the observable period during which the object is above an altitude of \SI{45}{\degree} is determined for a declination of \ang{-5}, which is typical for variable objects found, for example, in the Orion nebula. The corresponding visibilities are then calculated using the UT configuration of the VLTI. An exemplary plot of visibility and closure phase for different hour angles, together with the baseline variation and uv coverage for different hour angles for the inclined model in the active state and with an inner radius of \SI{0.5}{\au}, is shown in Fig.~\ref{fig:blvar_eg}.

It can be seen that the changes in the longer baselines are up to 30\% depending on the hour angle. This corresponds to changes larger than 0.1 in the observed visibilities even for timescales of around two hours. The measured closure phases also differ significantly depending on the triplet of telescopes used for the observation.

\begin{figure}[!ht]
    \centering
    \includegraphics[width=.8\linewidth]{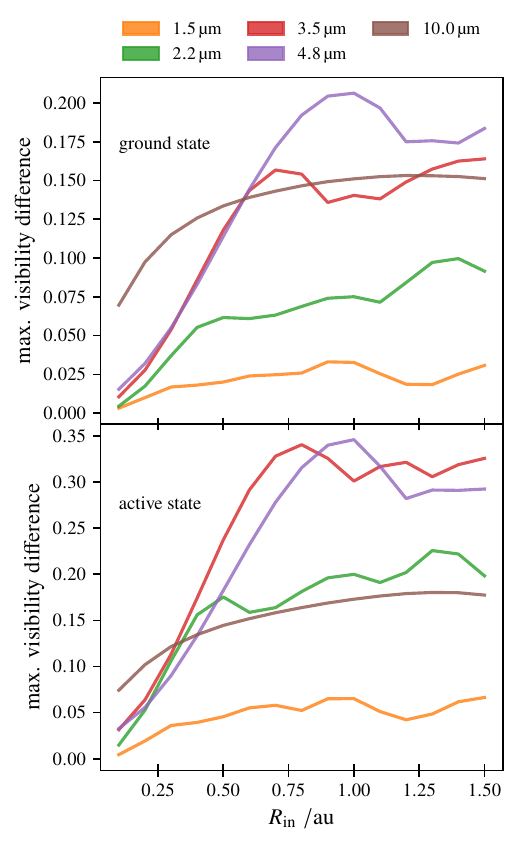}
    \caption{Maximal difference of visibility due to a change in hour angle of the models for the ground state (top) and the active state (bottom) with an inclination of \ang{45} as a function of the dust sublimation radius discussed in Sect.~\ref{subsec:SubRad} for the $H$, $K$, $L$, $M$, and $N$ bands.}
    \label{fig:blvar}
\end{figure}

\begin{figure}[!ht]
    \centering
    \includegraphics[width=.8\linewidth]{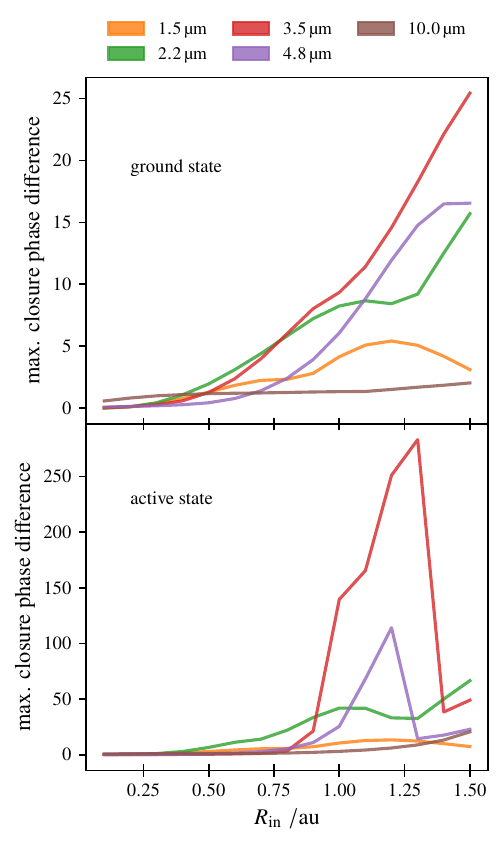}
    \caption{Same as Fig.~\ref{fig:blvar}, but for closure phases (see Sect.~\ref{subsec:basvar} for details).}
    \label{fig:blvar_cp}
\end{figure}

To quantify the impact of the baseline variation for all models and wavelengths, this procedure is repeated for all inclined models and the maximum differences of visibilities and closure phases are determined.
The corresponding plot of the maximum difference of the visibilities caused by a change in the hour angle for each model and wavelength is shown in Fig.~\ref{fig:blvar}.
The differences for all wavelengths except the $H$ and $K$ bands for both stellar effective temperatures are on the order of the changes discussed in Sect.~\ref{sec:Res}. The largest differences in the visibilities can be found for the models with the largest inner radii, that is the models for which the inner disk structures are best resolved. The same effect can be seen when taking the maximum differences of the closure phase into account. The resulting plot of the closure phases is shown in Fig.~\ref{fig:blvar_cp}. 
Since asymmetries are most prominent for the models with larger inner radii, the differences in the closure phase are also largest for these models. In the case of the $L$- and $M$-band simulations, some structures are resolved, leading to flips of the closure phase of \ang{360}.

The impact of baseline variation due to different hour angles is largely dependent on the brightness distribution. Thus, we provide an interactive online tool, \texttt{VLTI B-VAR}.\footnote{\url{https://vltib-var.streamlit.app}} It allows plots similar to Fig.~\ref{fig:blvar_eg} to be calculated for the model discussed in this section, as well as a Gaussian disk model for two disk inclinations. It is furthermore possible to upload and use customized intensity maps.

In conclusion, changes in the visibilities and closure phase due to baseline variation can be on the same order as differences due to a change in the disk illumination. It should therefore be checked that observations are done at similar hour angles to ensure comparability and to avoid false detections of variability.

\subsection{Observational timescales}\label{subsec:obscale}
In the previous section it was shown that observations of the same object should ideally be made at the same hour angle. However, this means that effects occurring on light-traveling timescales in the disk cannot be traced. For the model discussed in this paper, this timescale would be on the order of  10 to 20 hours. Additionally, the regions contributing most to the observables are located up to the $\tau = 10$ plane of the disk. The diffusive timescale of the radiation for these regions is of a factor of 1.25 to 2.5 larger than the light-traveling time. This means that the dust in the relevant regions of the disk is close to or at a radiative equilibrium state. The assumption of radiative equilibrium made for the simulations in Sect.~\ref{sec:Res} is thus well justified. A derivation of the respective timescales can be found in Appendix~\ref{sec:difftime}.

\section{Summary}\label{sec:sum}
We investigated the impact of a change in the brightness distribution of a circumstellar disk on the visibility and closure phase of VLTI observations and the corresponding changes in the $V$- and $N$-band photometry. This was done by applying radiative transfer simulations to a basic, often applied disk model. We isolated the effects of a different illumination of the disk due to geometrical changes from changes due to an increase in stellar luminosity. Finally, we studied the combination of the two mechanisms, that is, the change in the inner radius of a disk due to dust sublimation caused by an increase in the stellar luminosity. In order to take the different stellar SED caused by accretion events into account, we performed all simulations for stellar effective temperatures of \SI{4000}{\K} (ground state) and \SI{8000}{\K} (active state), respectively.

We found that changing the inner radius of the disk model while keeping the stellar luminosity constant leads to measurable changes in the visibilities and closure phases of longer wavelengths, while the impact on the photometric flux is small.
Increasing the stellar luminosity with an inner radius of the disk fixed outside the sublimation radius, we found that changes in the visibilities and closure phases for the larger wavelengths are even stronger. While the $V$-band photometry follows the increase in the stellar luminosity directly, the $N$-band photometric flux is always lower, but still increasing on a similar order proportional to the stellar luminosity.
Combining a change in the inner radius of the disk via dust sublimation due to an increased luminosity, we find significant differences in the visibilities and closure phases at all involved wavelengths. The reason for this is an increase in the star-to-disk brightness ratio over all wavelengths, which is caused by the temperature distribution of the dust now covering also the higher temperatures up to the sublimation temperature. This effect is even stronger for the model of the active state. In this case, some visibilities calculated for this specific model are even close to zero, hinting at well-resolved structures caused by the change in the disk illumination. 
The $V$- and $N$-band photometric fluxes are similar to the simulation with an increase in the stellar luminosity only, which is expected when comparing to the simulations with constant stellar luminosity.

Finally, we discussed the resulting consequences for the planning of observations. In this context special attention should be paid to the hour angle at which the observations are performed to avoid deviations of visibility and closure phase due to baseline variation when comparing observations of different epochs.

In summary, we found that interferometric observations can provide valuable insights into the brightness structure of variable circumstellar disks. Since it is possible to refer changes in the $V$- and $N$-band photometry to changes in the interferometric observables of the $H$, $K$, $L$, $M$, and $N$ bands, existing photometric variability measurements could be used to identify promising targets for future multi-epoch interferometric observations.

We find that the level of photometric variability corresponding to measurable changes in the interferometric observables derived in our study ($>\SI{0.1}{mag}$, i.e., a factor of 2) is comparable to that of median short-term photometric variability of low-mass TTauri stars \citep[see, e.g.,][]{Pouilly2021,Cody2022}.
Additionally, outbursts in young stellar objects also show an increase in brightness of $\SI{2.5}{mag}-\SI{6}{mag}$ (factors of several tens to hundreds) for long-term outburst events \citep[see][]{Fischer2022}.
Thus, the results of this paper encourage the investigation of these objects using interferometry.

\bibliographystyle{aa} 
\bibliography{lit} 

\begin{thebibliography}{32}
\expandafter\ifx\csname natexlab\endcsname\relax\def\natexlab#1{#1}\fi

\bibitem[{{Andrews} {et~al.}(2010){Andrews}, {Wilner}, {Hughes}, {Qi}, \&
  {Dullemond}}]{Andrews2010}
{Andrews}, S.~M., {Wilner}, D.~J., {Hughes}, A.~M., {Qi}, C., \& {Dullemond},
  C.~P. 2010, \apj, 723, 1241

\bibitem[{{Bjorkman} \& {Wood}(2001)}]{BjorkmanWood2001}
{Bjorkman}, J.~E. \& {Wood}, K. 2001, \apj, 554, 615

\bibitem[{{Brunngr{\"a}ber} \& {Wolf}(2018)}]{Brunngraeber2018}
{Brunngr{\"a}ber}, R. \& {Wolf}, S. 2018, \aap, 611, A90

\bibitem[{{Brunngr{\"a}ber} {et~al.}(2016){Brunngr{\"a}ber}, {Wolf}, {Ratzka},
  \& {Ober}}]{Brunngraeber2016}
{Brunngr{\"a}ber}, R., {Wolf}, S., {Ratzka}, T., \& {Ober}, F. 2016, \aap, 585,
  A100

\bibitem[{{Chen} {et~al.}(2018){Chen}, {K{\'o}sp{\'a}l}, {{\'A}brah{\'a}m},
  {Kreplin}, {Matter}, \& {Weigelt}}]{Chen2018}
{Chen}, L., {K{\'o}sp{\'a}l}, {\'A}., {{\'A}brah{\'a}m}, P., {et~al.} 2018,
  \aap, 609, A45

\bibitem[{{Chen} {et~al.}(2019){Chen}, {Mo{\'o}r}, {Kreplin}, {K{\'o}sp{\'a}l},
  {{\'A}brah{\'a}m}, {Matter}, {Carmona}, {Hofmann}, {Schertl}, \&
  {Weigelt}}]{Chen2019}
{Chen}, L., {Mo{\'o}r}, A., {Kreplin}, A., {et~al.} 2019, \apjl, 887, L32

\bibitem[{{Cody} {et~al.}(2022){Cody}, {Hillenbrand}, \& {Rebull}}]{Cody2022}
{Cody}, A.~M., {Hillenbrand}, L.~A., \& {Rebull}, L.~M. 2022, \aj, 163, 212

\bibitem[{{Davies} {et~al.}(2020){Davies}, {Kraus}, {Harries}, {Monnier},
  {Kloppenborg}, {Aarnio}, {Baron}, {Lopez}, {Millan-Gabet}, {Parks},
  {Pedretti}, {Perraut}, {Sturmann}, {Sturmann}, {ten Brummelaar}, \&
  {Touhami}}]{Davies2020}
{Davies}, C.~L., {Kraus}, S., {Harries}, T.~J., {et~al.} 2020, \apj, 897, 31

\bibitem[{{Draine} \& {Lee}(1984)}]{Draine1984}
{Draine}, B.~T. \& {Lee}, H.~M. 1984, \apj, 285, 89

\bibitem[{{Draine} \& {Malhotra}(1993)}]{DraineMalhotra1993}
{Draine}, B.~T. \& {Malhotra}, S. 1993, \apj, 414, 632

\bibitem[{{Fischer} {et~al.}(2022){Fischer}, {Hillenbrand}, {Herczeg},
  {Johnstone}, {K{\'o}sp{\'a}l}, \& {Dunham}}]{Fischer2022}
{Fischer}, W.~J., {Hillenbrand}, L.~A., {Herczeg}, G.~J., {et~al.} 2022, arXiv
  e-prints, arXiv:2203.11257

\bibitem[{{Flock} {et~al.}(2016){Flock}, {Fromang}, {Turner}, \&
  {Benisty}}]{Flock2016}
{Flock}, M., {Fromang}, S., {Turner}, N.~J., \& {Benisty}, M. 2016, \apj, 827,
  144

\bibitem[{{Hofmann} {et~al.}(2022){Hofmann}, {Bensberg}, {Schertl}, {Weigelt},
  {Wolf}, {Meilland}, {Millour}, {Waters}, {Kraus}, {Ohnaka}, {Lopez},
  {Petrov}, {Lagarde}, {Berio}, {Allouche}, {Robbe-Dubois}, {Jaffe}, {Henning},
  {Paladini}, {Sch{\"o}ller}, {M{\'e}rand}, {Glindemann}, {Beckmann},
  {Heininger}, {Bettonvil}, {Zins}, {Woillez}, {Bristow}, {Stee}, {Vakili},
  {van Boekel}, {Hogerheijde}, {Dominik}, {Augereau}, {Matter}, {Hron},
  {Pantin}, {Rivinius}, {de Wit}, {Varga}, {Klarmann}, {Meisenheimer},
  {G{\'a}mez Rosas}, {Burtscher}, {Leftley}, {Isbell}, {Yoffe}, {Kokoulina},
  {Danchi}, {Cruzal{\`e}bes}, {Domiciano de Souza}, {Drevon}, {Hocd{\'e}},
  {Kreplin}, {Labadie}, {Connot}, {Nu{\ss}baum}, {Lehmitz}, {Antonelli},
  {Graser}, \& {Leinert}}]{Hofmann2022}
{Hofmann}, K.~H., {Bensberg}, A., {Schertl}, D., {et~al.} 2022, \aap, 658, A81

\bibitem[{{Isella} \& {Natta}(2005)}]{Isella2005}
{Isella}, A. \& {Natta}, A. 2005, \aap, 438, 899

\bibitem[{{Joy}(1945)}]{Joy1945}
{Joy}, A.~H. 1945, \apj, 102, 168

\bibitem[{{Kluska} {et~al.}(2016){Kluska}, {Benisty}, {Soulez}, {Berger}, {Le
  Bouquin}, {Malbet}, {Lazareff}, \& {Thi{\'e}baut}}]{Kluska2016}
{Kluska}, J., {Benisty}, M., {Soulez}, F., {et~al.} 2016, \aap, 591, A82

\bibitem[{{Kobus} {et~al.}(2020){Kobus}, {Wolf}, {Ratzka}, \&
  {Brunngr{\"a}ber}}]{Kobus2020}
{Kobus}, J., {Wolf}, S., {Ratzka}, T., \& {Brunngr{\"a}ber}, R. 2020, \aap,
  642, A104

\bibitem[{{Laor} \& {Draine}(1993)}]{Laor1993}
{Laor}, A. \& {Draine}, B.~T. 1993, \apj, 402, 441

\bibitem[{{Lopez} {et~al.}(2022){Lopez}, {Lagarde}, {Petrov}, {Jaffe},
  {Antonelli}, {Allouche}, {Berio}, {Matter}, {Meilland}, {Millour},
  {Robbe-Dubois}, {Henning}, {Weigelt}, {Glindemann}, {Agocs}, {Bailet},
  {Beckmann}, {Bettonvil}, {van Boekel}, {Bourget}, {Bresson}, {Bristow},
  {Cruzal{\`e}bes}, {Eldswijk}, {Fante{\"\i} Caujolle}, {Gonz{\'a}lez Herrera},
  {Graser}, {Guajardo}, {Heininger}, {Hofmann}, {Kroes}, {Laun}, {Lehmitz},
  {Leinert}, {Meisenheimer}, {Morel}, {Neumann}, {Paladini}, {Percheron},
  {Riquelme}, {Schoeller}, {Stee}, {Venema}, {Woillez}, {Zins},
  {{\'A}brah{\'a}m}, {Abadie}, {Abuter}, {Accardo}, {Adler}, {Alonso},
  {Augereau}, {B{\"o}hm}, {Bazin}, {Beltran}, {Bensberg}, {Boland}, {Brast},
  {Burtscher}, {Castillo}, {Chelli}, {Cid}, {Clausse}, {Connot}, {Conzelmann},
  {Danchi}, {Delbo}, {Drevon}, {Dominik}, {van Duin}, {Ebert}, {Eisenhauer},
  {Flament}, {Frahm}, {G{\'a}mez Rosas}, {Gabasch}, {Gallenne}, {Garces},
  {Girard}, {Glazenborg}, {Gont{\'e}}, {Guitton}, {de Haan}, {Hanenburg},
  {Haubois}, {Hocd{\'e}}, {Hogerheijde}, {ter Horst}, {Hron}, {Hummel},
  {Hubin}, {Huerta}, {Idserda}, {Isbell}, {Ives}, {Jakob}, {Jask{\'o}},
  {Jochum}, {Klarmann}, {Klein}, {Kragt}, {Kuindersma}, {Kokoulina}, {Labadie},
  {Lacour}, {Leftley}, {Le Poole}, {Lizon}, {Lopez}, {Lykou}, {M{\'e}rand},
  {Marcotto}, {Mauclert}, {Maurer}, {Mehrgan}, {Meisner}, {Meixner}, {Mellein},
  {Menut}, {Mohr}, {Mosoni}, {Navarro}, {Nu{\ss}baum}, {Pallanca}, {Pantin},
  {Pasquini}, {Phan Duc}, {Pott}, {Pozna}, {Richichi}, {Ridinger}, {Rigal},
  {Rivinius}, {Roelfsema}, {Rohloff}, {Rousseau}, {Salabert}, {Schertl},
  {Schuhler}, {Schuil}, {Shabun}, {Soulain}, {Stephan}, {Toledo}, {Tristram},
  {Tromp}, {Vakili}, {Varga}, {Vinther}, {Waters}, {Wittkowski}, {Wolf},
  {Wrhel}, \& {Yoffe}}]{Lopez2022}
{Lopez}, B., {Lagarde}, S., {Petrov}, R.~G., {et~al.} 2022, \aap, 659, A192

\bibitem[{{Lucy}(1999)}]{Lucy1999}
{Lucy}, L.~B. 1999, \aap, 344, 282

\bibitem[{{Mathis} {et~al.}(1977){Mathis}, {Rumpl}, \&
  {Nordsieck}}]{MathisRumplNordsieck1977}
{Mathis}, J.~S., {Rumpl}, W., \& {Nordsieck}, K.~H. 1977, The Astrophysical
  Journal, 217, 425

\bibitem[{{Poppenhaeger} {et~al.}(2015){Poppenhaeger}, {Cody}, {Covey},
  {G{\"u}nther}, {Hillenbrand}, {Plavchan}, {Rebull}, {Stauffer}, {Wolk},
  {Espaillat}, {Forbrich}, {Gutermuth}, {Hora}, {Morales-Calder{\'o}n}, \&
  {Song}}]{Poppenhaeger2015}
{Poppenhaeger}, K., {Cody}, A.~M., {Covey}, K.~R., {et~al.} 2015, \aj, 150, 118

\bibitem[{{Pouilly} {et~al.}(2021){Pouilly}, {Bouvier}, {Alecian}, {Alencar},
  {Cody}, {Donati}, {Grankin}, {Rebull}, \& {Folsom}}]{Pouilly2021}
{Pouilly}, K., {Bouvier}, J., {Alecian}, E., {et~al.} 2021, \aap, 656, A50

\bibitem[{{Rebull} {et~al.}(2015){Rebull}, {Stauffer}, {Cody}, {G{\"u}nther},
  {Hillenbrand}, {Poppenhaeger}, {Wolk}, {Hora}, {Hernandez}, {Bayo}, {Covey},
  {Forbrich}, {Gutermuth}, {Morales-Calder{\'o}n}, {Plavchan}, {Song}, {Bouy},
  {Terebey}, {Cuillandre}, \& {Allen}}]{Rebull2015}
{Rebull}, L.~M., {Stauffer}, J.~R., {Cody}, A.~M., {et~al.} 2015, \aj, 150, 175

\bibitem[{{Reissl} {et~al.}(2016){Reissl}, {Wolf}, \& {Brauer}}]{Reissl2016}
{Reissl}, S., {Wolf}, S., \& {Brauer}, R. 2016, Astronomy $\&$ Astrophysics,
  593, A87

\bibitem[{{Sauter} {et~al.}(2009){Sauter}, {Wolf}, {Launhardt}, {Padgett},
  {Stapelfeldt}, {Pinte}, {Duch{\^e}ne}, {M{\'e}nard}, {McCabe}, {Pontoppidan},
  {Dunham}, {Bourke}, \& {Chen}}]{Sauter2009}
{Sauter}, J., {Wolf}, S., {Launhardt}, R., {et~al.} 2009, \aap, 505, 1167

\bibitem[{{Schegerer} {et~al.}(2009){Schegerer}, {Wolf}, {Hummel}, {Quanz}, \&
  {Richichi}}]{Schegerer2009}
{Schegerer}, A.~A., {Wolf}, S., {Hummel}, C.~A., {Quanz}, S.~P., \& {Richichi},
  A. 2009, \aap, 502, 367

\bibitem[{{Shakura} \& {Sunyaev}(1973)}]{ShakuraSunjaev}
{Shakura}, N.~I. \& {Sunyaev}, R.~A. 1973, Astronomy $\&$ Astrophysics, 24, 337

\bibitem[{{Stauffer} {et~al.}(2016){Stauffer}, {Cody}, {Rebull}, {Hillenbrand},
  {Turner}, {Carpenter}, {Carey}, {Terebey}, {Morales-Calder{\'o}n}, {Alencar},
  {McGinnis}, {Sousa}, {Bouvier}, {Venuti}, {Hartmann}, {Calvet}, {Micela},
  {Flaccomio}, {Song}, {Gutermuth}, {Barrado}, {Vrba}, {Covey}, {Herbst},
  {Gillen}, {Medeiros Guimar{\~a}es}, {Bouy}, \& {Favata}}]{Stauffer2016}
{Stauffer}, J., {Cody}, A.~M., {Rebull}, L., {et~al.} 2016, \aj, 151, 60

\bibitem[{{Weingartner} \& {Draine}(2001)}]{Weingartner2001}
{Weingartner}, J.~C. \& {Draine}, B.~T. 2001, \apj, 548, 296

\bibitem[{{Woitke} {et~al.}(2019){Woitke}, {Kamp}, {Antonellini}, {Anthonioz},
  {Baldovin-Saveedra}, {Carmona}, {Dionatos}, {Dominik}, {Greaves},
  {G{\"u}del}, {Ilee}, {Liebhardt}, {Menard}, {Min}, {Pinte}, {Rab}, {Rigon},
  {Thi}, {Thureau}, \& {Waters}}]{Woitke2019}
{Woitke}, P., {Kamp}, I., {Antonellini}, S., {et~al.} 2019, \pasp, 131, 064301

\bibitem[{{Wolk} {et~al.}(2018){Wolk}, {G{\"u}nther}, {Poppenhaeger},
  {Winston}, {Rebull}, {Stauffer}, {Gutermuth}, {Cody}, {Hillenbrand},
  {Plavchan}, {Covey}, \& {Song}}]{Wolk2018}
{Wolk}, S.~J., {G{\"u}nther}, H.~M., {Poppenhaeger}, K., {et~al.} 2018, \aj,
  155, 99

\end{thebibliography}

\begin{appendix}

\section{Diffusion timescales}\label{sec:difftime}
In the case of high optical depth, the radiative transfer of energy $u$ in a medium can be described using a diffusion approximation following Fick's law of diffusion with a diffusion constant $D$:
\begin{equation}
    \label{eq:diff}
    \frac{du}{dt} = -D \frac{d^2u}{dx^2}.
\end{equation}
The distribution of the energy in the medium can be described using the ansatz of a Gauss function:
\begin{equation}
    \label{eq:gauss}
    u(x,t) = \frac{1}{\sqrt{4\pi D t}} \exp\left(-\frac{x^2}{Dt}\right).
\end{equation}
We now define the diffusion timescale $T_{2\sigma}$ as the time it takes to distribute $2\sigma$ of the energy over a distance $L$:
\begin{equation}
    \label{eq:difftime}
    2 \sigma = 2 \sqrt{2Dt} = L \Rightarrow \fbox{$T_{2\sigma} = \frac{L^2}{8 D}$}.
\end{equation}
By applying $D= \frac{c}{\kappa}$ with speed of light $c$ and absorption coefficient $\kappa$, we find a relation between the light traveling time $T_l = L/c$ and the diffusion timescale:
\begin{equation}
    \label{eq:ltodiff}
    \frac{T_{2\sigma}}{T_l} = \frac{L^2/8D}{L/c} \Rightarrow \fbox{$T_{2\sigma} \approx T_l \frac{\tau}{8}$}.
\end{equation}
It should be noted that this only holds true for high optical depth, that is $\frac{\tau}{8} >> 1$.
In case of the model disk discussed in Sect.~\ref{sec:Meth}, the optical depth for all wavelengths is larger than $10^3$. Thus, only the upper dust layers of the disk are effected on light-traveling timescales. Since the contribution of these layers to the visibilities and closure phases are comparably small, we can assume stationary solutions for the effects discussed in Sect.~\ref{sec:Res}.

\section{Modified inner rim}\label{sec:ModRin}
In Sect.~\ref{subsec:SubRad} we present simulations of the model described in Sect.~\ref{sec:Meth} for increasing inner radii due to sublimation caused by an increasing stellar luminosity. To test the influence of a more complex inner rim shape on the discussed trends, we repeated the simulations of the inclined disk for a model with a modified inner rim. Following \cite{Hofmann2022}, we used a simple parameterization of different curved inner rim shapes using a modification of the scale height given in Eq.~\ref{eq:scale_height}
\begin{equation}
    \label{eq:modrin}
    h(r) = h_0(R_{\varepsilon}) \cdot \left(\frac{r - R_{\rm in}}{R_{\varepsilon}-R_{\rm in}}\right)^\varepsilon ,
\end{equation}
with the shape parameter $\varepsilon$ and the radius $R_{\varepsilon}$ up to which the rim is modified.

\begin{figure}[!ht]
    \centering
    \includegraphics[width=1.\linewidth]{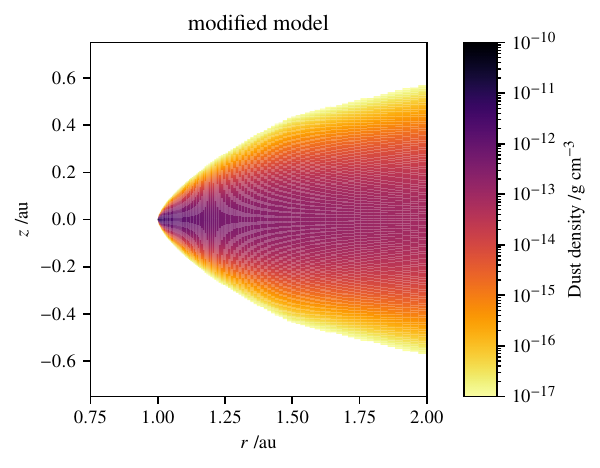}
    \caption{Vertical cut through the density distribution of the model with $R_{\rm in} = \SI{1}{\au}$ and modified inner rim. The parameters of the modified inner rim are $R_{\varepsilon} = 1.5 R_{\rm in}$ and $\varepsilon = 0.7$ (see Appendix~\ref{sec:ModRin} for details).}
    \label{fig:mod_dens}
\end{figure}

\begin{figure*}[!ht]
    \centering
    \includegraphics[width=1.\linewidth]{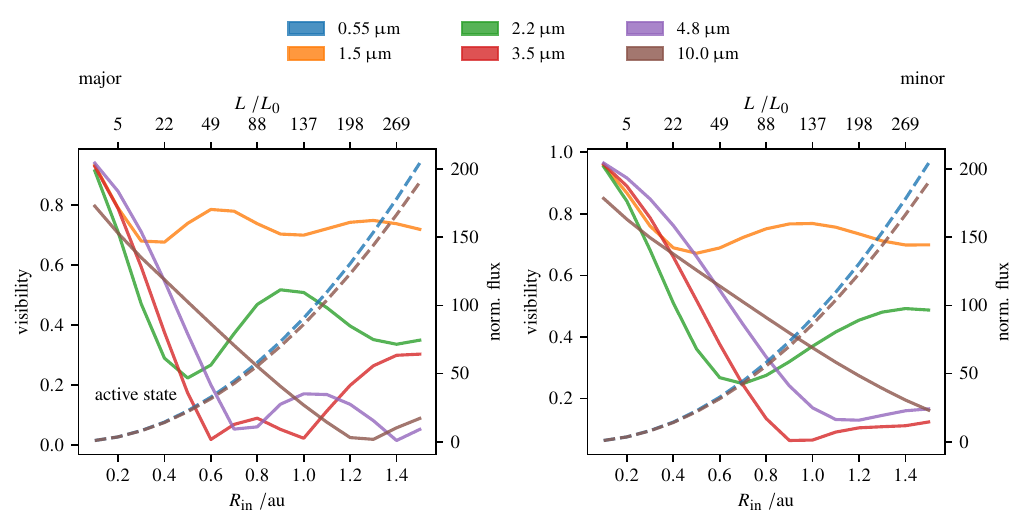}
    \caption{Visibilities of models with modified inner rim for the active state with an inclination of \ang{45} as a function of  stellar luminosity and the corresponding sublimation radius for the $H$, $K$, $L$, $M$, and $N$ bands (solid; left axis) for baseline orientations along the major (left) and minor axis (right) of the disk image. The corresponding fluxes in the photometric $V$ and $N$ bands are also shown (dashed; right axis). The $V$- and $N$-band fluxes are normalized with respect to the reference model with $L = L_0$. The stellar luminosity of each model is increased such that the sublimation radius matches the given $R_{\textrm{in}}$ (see Sect.~\ref{subsec:SubRad} for details).}
    \label{fig:subratio45_mod}
\end{figure*}

We chose parameters of $R_{\varepsilon} = 1.5 R_{\rm in}$ and $\varepsilon = 0.7$, which correspond to inner rim shapes discussed for example by \cite{Isella2005} or \cite{Davies2020}.
A vertical cut of the resulting density distribution of the model with $R_{\rm in} = \SI{1}{\au}$ is shown in Fig.~\ref{fig:mod_dens}. The corresponding visibilities of the simulations for the active state as well as the photometric fluxes of the $V$ and $N$ bands can be found in Fig.~\ref{fig:subratio45_mod}.
The qualitative trends identified in Fig.~\ref{fig:subratio45} in Sect.~\ref{subsec:SubRad} are also found for this modified model. However, the minima of the visibilities are shifted to smaller inner radii since the bright emission of the inner rim is now distributed over a larger area and gets resolved at smaller inner radii.
While the overall order of magnitude of the changes in the visibilities for the baseline orientation along the minor axis are similar to those found in the simulations for the model with the unmodified inner rim, all visibilities are significantly lower for the baseline orientation along the major axis. This is caused by the brightness of the inner rim that is now smeared out over a larger area of the brightness distribution, contributing to the resolved brightness distribution.

\begin{figure}[!ht]
    \centering
    \includegraphics[width=1.\linewidth]{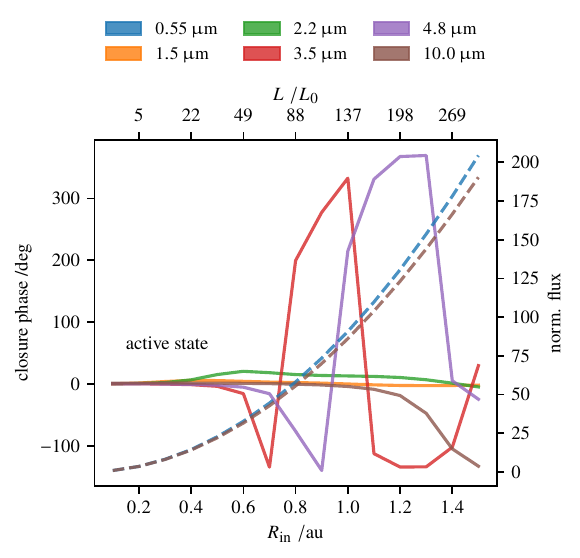}
    \caption{Closure phases of models with modified inner rim for the active state with an inclination of \ang{45} as a function of  stellar luminosity  and the corresponding sublimation radius for the $H$, $K$, $L$, $M$, and $N$ bands (solid; left axis). The corresponding fluxes in the photometric $V$ and $N$ bands are also shown (dashed; right axis). The $V$- and $N$-band fluxes are normalized with respect to the reference model with $L = L_0$. The stellar luminosity of each model is increased such that the sublimation radius matches the given $R_{\textrm{in}}$ (see Sect.~\ref{subsec:SubRad} for details).}
    \label{fig:subratiocp_modrin}
\end{figure}

The corresponding closure phases calculated in analogy to Fig.~\ref{fig:subratiocp} for the model with the modified inner rim in the active state are shown in Fig.~\ref{fig:subratiocp_modrin}.
Again, we find that the minima and maxima of the closure phases are shifted toward shorter inner radii. The flips in the sign of the closure phase for the $L$ and $M$ bands also correspond to the respective minima of the visibilities seen in Fig.~\ref{fig:subratio45_mod}. In turn, these effects can be traced back to the spatially larger brightness structure of the modified inner rim.

In summary, a change in the inner rim geometry is not significantly changing the trends found in Sect.~\ref{sec:Res}.
\end{appendix}

\end{document}